\documentclass[aps,prd,twocolumn,superscriptaddress,amsmath,amssymb,nofootinbib,eqsecnum,floatfix]{revtex4-1}
\pdfoutput=1
\usepackage{comment}
\usepackage{graphicx}
\usepackage[colorlinks=true,allcolors={blue}]{hyperref}
\usepackage{color}
\usepackage{fancyhdr}

\fancypagestyle{firstpage}{\fancyhf{}\cfoot{\thepage}}
\let\ab\allowbreak

\DeclareMathOperator{\arcsinh}{arcsinh}
\DeclareMathOperator{\arccosh}{arccosh}
\DeclareMathOperator{\sn}{sn}
\DeclareMathOperator{\cn}{cn}
\DeclareMathOperator{\dn}{dn}
\DeclareMathOperator{\fn}{fn}
\newcommand{\munu}{{\mu\nu}}

\begin{document}

\title{Exact Geodesic Distances in FLRW Spacetimes}

\author{William J.\ Cunningham}
\affiliation{Department of Physics, Northeastern University, 360 Huntington Ave. Boston, MA 02115, United States}
\email{w.cunningham@northeastern.edu}

\author{David Rideout}
\affiliation{Department of Mathematics, University of California, San Diego, 9500 Gilman Dr. Mail Code 0112, La Jolla, CA 92093, United States}

\author{James Halverson}
\affiliation{Department of Physics, Northeastern University, 360 Huntington Ave. Boston, MA 02115, United States}

\author{Dmitri Krioukov}
\affiliation{Department of Physics, Department of Mathematics, Department of Electrical \& Computer Engineering, Northeastern University, 360 Huntington Ave. Boston, MA 02115, United States}
\email{dima@northeastern.edu}

\date{\today}

\begin{abstract}
  Geodesics are used in a wide array of applications in cosmology and
  astrophysics. However, it is not a trivial task to efficiently calculate
  exact geodesic distances in an arbitrary spacetime. We show that in
  spatially flat $(3+1)$-dimensional Friedmann-Lema\^itre-Robertson-Walker (FLRW)
  spacetimes, it is possible to integrate the second-order geodesic
  differential equations, and derive a general method for finding both
  timelike and spacelike distances given initial-value or boundary-value
  constraints. In flat spacetimes with either dark energy or matter, whether
  dust, radiation, or a stiff fluid, we find an exact closed-form solution for
  geodesic distances. In spacetimes with a mixture of dark energy and matter,
  including spacetimes used to model our physical universe, there exists no
  closed-form solution, but we provide a fast numerical method to compute geodesics.
  A general method is also described for determining the geodesic
  connectedness of an FLRW manifold, provided only its scale factor.
\end{abstract}

\maketitle

\section{Introduction}
Cosmic microwave background experiments such as COBE~\cite{ref:smoot1992},
WMAP~\cite{ref:hinshaw2013}, and Planck~\cite{ref:ade2016} provide evidence for
both early time cosmic inflation~\cite{ref:guth1981,ref:linde1982} and late time
acceleration~\cite{perlmutter1998discovery,riess1998observational}, with interesting
dynamics in between explaining many features of the universe, many of which are
remarkably accurately predicted by the $\Lambda$CDM model~\cite{ref:peebles1982,%
ref:turner1984,ref:blumenthal1984,ref:davis1985}. These and other experiments in
recent decades have demonstrated that, to a high degree of precision, at large
scales the visible universe is spatially homogeneous, isotropic, and flat, i.e.,
that its spacetime is described by the Friedmann-Lema\^itre-Robertson-Walker (FLRW)
metric. FLRW spacetimes are therefore of particular interest in modern cosmology. \par

Here we develop a method for the exact calculation of the geodesic
distance between any given pair of events in any flat FLRW spacetime.
Geodesics and geodesic distances naturally arise
in a wide variety of investigations not only in cosmology, but also in
astrophysics and quantum gravity, with topics ranging from the horizon and dark energy
problems, to gravitational lensing,
to evaluating the observational signatures of cosmic bubble collisions and modified gravity theories,
to the AdS/CFT correspondence~\cite{ellis1999deviation,ref:gron2007,albareti2012focusing,demianski2003approximate,hirata2005superhorizon,bikwa2012photon,doplicher2013quantum,melia2013proper,bhattacharya2017cosmic,pyne1996lens,park2008rigorous,sereno2009role,mukohyama2009dark,mukohyama2009caustic,traschen1986large,futamase1989light,cooperstock1998influence,caldwell2001shortcuts,dappiaggi2008stable,kaloper2010mcvitties,melia2013ct,bahrami2017saturating,koyama2001strongly,dong2012frw,dong2012aspects,minton2008new,ref:wainwright2014,ref:hagala2016}.
Closed-form solutions of the geodesic equations are also quite useful in validating a particular FLRW model by investigating how curvature, quintessence, local shear terms, etc., affect observational data.  These solutions are of perhaps the greatest and most direct utility in large-scale N-body simulations, e.g., studying the large scale structure formation, which can benefit greatly from using such solutions by avoiding the costly numerical integration of the geodesic differential equations~\cite{ref:adamek2016,ref:koksbang2015,ref:bibiano2017}. \par

For a general spacetime, solving the geodesic
equations exactly for given initial-value or boundary-value constraints is
intractable, although it may be possible in some cases. For example,
in $(3+1)$-dimensional de Sitter space, which represents a spacetime with only dark energy
and is a maximally symmetric solution to Einstein's equations, it turns out to be rather simple to study geodesics by embedding the manifold into flat $(4+1)$-dimensional Minkowski space $\mathbb{M}^5$. This construction was originally realized by de Sitter himself~\cite{ref:desitter1917}, and was later studied by Schr\"odinger~\cite{ref:schrodinger1956}.
In Sec.~\ref{sec:desitter} we review how geodesics may be found using the unique geometric properties of this manifold. \par

However, it is not so easy to explicitly calculate geodesic distances in other
FLRW spacetimes except under certain assumptions. One approach would be to
follow de Sitter's philosophy by finding an embedding into a
higher-dimensional manifold. Such an embedding always exists due to the
Campbell-Magaard theorem, which states that any analytic $n$-dimensional
Riemannian manifold may be locally embedded into an $(n+1)$-dimensional
Ricci-flat space~\cite{ref:campbell1926,ref:magaard1963}, combined with a
theorem due to A.~Friedman extending the result to pseudo-Riemannian
manifolds~\cite{ref:friedman1961,ref:friedman1965}. In fact, the embedding map
is given explicitly by J.~Rosen in~\cite{ref:rosen1965}. However, it has since been shown
that the metric in the embedding space is block diagonal with respect to the
embedded surface, i.e., when the geodesic is constrained to the
$(3+1)$-dimensional subspace we regain the original $(3+1)$-dimensional
geodesic differential equations and we learn nothing
new~\cite{ref:romero1996}. \par

Instead, in Sec.~\ref{sec:exact_solutions} we solve directly the geodesic differential equations for a general FLRW spacetime in terms of the scale factor and a set of initial-value or boundary-value constraints. The final geodesic distance can be written as an integral which is a function of the boundary conditions and one extra constant $\mu$, defined by a transcendental integral equation. This constant proves to be useful in a number of ways: it tells us if a manifold is geodesically connected provided only the scale factor. The solution of the integral equation defining this constant exists only if a geodesic exists for a given set of boundary conditions, and it helps one to find the geodesic distance, if such a geodesic exists. \par

We then give some examples in Sec.~\ref{sec:examples} to show that for many scale factors of interest, we can find a closed-form solution. In cases where no closed-form solution exists, we can still transform the problem into one which is suitable for fast numerical integration. Concluding remarks are in Sec.~\ref{sec:conclusion}.

\section{Review of FLRW Spacetimes and de Sitter Embeddings}
\label{sec:frw_spacetimes}
Friedmann-Lema\^itre-Robertson-Walker (FLRW) spacetimes are spatially homogeneous and isotropic $(3+1)$-dimensional Lorentzian manifolds which are solutions to Einstein's equations~\cite{ref:griffiths2009}. These manifolds have a metric $g_\munu$ with $\mu$, $\nu \in \{0,1,2,3\}$ that, when diagonalized in a given coordinate system, gives an invariant interval $ds^2=g_\munu\,dx^\mu\,dx^\nu$ of the form
\begin{equation}
\label{eq:invariant_interval}
ds^2=-dt^2+a(t)^2\,d\Sigma^2\,,
\end{equation}
where $a(t)$ is the scale factor, which describes how space expands with time $t$, and $d\Sigma$ is the spatial 
metric given by $d\Sigma^2=dr^2+r^2(d\theta^2+\sin^2\theta\,d\phi^2)$ for flat space in spherical coordinates $(r,\theta,\phi)$ that we use hereafter. The scale factor is found by solving Friedmann's equation, the differential equation given by the $\mu=\nu=0$ component of Einstein's equations:
\begin{equation}
\label{eq:friedmann}
\left(\frac{\dot a}{a}\right)^2 = \frac \Lambda 3 + \frac{c}{a^{3g}}\,,
\end{equation}
where $\Lambda$ is the cosmological constant, $g$ parametrizes the type of matter within the spacetime,
$c$ is a constant proportional to the matter density, and we have assumed spatial flatness in our choice of $d\Sigma$. The scale factors for manifolds which represent spacetimes with dark energy ($\Lambda$), dust ($D$), radiation ($R$), a stiff fluid ($S$),%
\footnote{Stiff fluids are exotic forms of matter which have a speed of sound equal to the speed of light. They have been studied in a variety of models of the early universe, including kination fields, self-interacting (warm) dark matter, and Ho\^rava-Lifshitz cosmologies~\cite{ref:dutta2010}.}
or some combination (e.g., $\Lambda D$ for dark energy and dust matter) are given by~\cite{ref:griffiths2009}
\begin{subequations}
\label{eq:scale_factors}
\begin{align}
a_\Lambda(t) &= \lambda e^{t/\lambda} \label{eq:scale_factor_lambda}\,, \\
a_D(t) &= \alpha\left(\frac{3t}{2\lambda}\right)^{2/3} \label{eq:scale_factor_D}\,, \\
a_R(t) &= \alpha^{3/4}\left(\frac{2t}{\lambda}\right)^{1/2} \label{eq:scale_factor_R}\,, \\
a_S(t) &= \alpha^{1/2}\left(\frac{3t}{\lambda}\right)^{1/3} \label{eq:scale_factor_S}\,, \\
a_{\Lambda D}(t) &= \alpha\sinh^{2/3}\left(\frac{3t}{2\lambda}\right) \label{eq:scale_factor_lambda_D}\,, \\
a_{\Lambda R}(t) &= \alpha^{3/4}\sinh^{1/2}\left(\frac{2t}{\lambda}\right) \label{eq:scale_factor_lambda_R}\,, \\
a_{\Lambda S}(t) &= \alpha^{1/2}\sinh^{1/3}\left(\frac{3t}{\lambda}\right) \label{eq:scale_factor_lambda_S}\,,
\end{align}
\end{subequations}
where $\lambda$ and $\alpha\equiv(c\lambda^2)^{1/3}$ are the temporal and spatial scale-setting parameters.
In manifolds with dark energy, i.e., $\Lambda>0$, $\lambda\equiv\sqrt{3/\Lambda}$.

\subsection{de Sitter Spacetime}
\label{sec:desitter}
The de Sitter spacetime is one of the first and best studied spacetimes: de Sitter himself recognized that the $(3+1)$-dimensional manifold d$\mathbb{S}^4$ can be visualized as a single-sheet hyperboloid embedded in $\mathbb{M}^5$, defined by
\begin{equation}
-z_0^2+z_1^2+z_2^2+z_3^2+z_4^2=\lambda^2\,,
\end{equation}
where $\lambda$ is the pseudo-radius of the hyperboloid~\cite{ref:desitter1917}. The injection $\chi\,:\,\mathrm{d}\mathbb{S}^4\hookrightarrow\mathbb{M}^5\,,\,\chi(x)\mapsto z$ is
\begin{align}
\begin{aligned}
\frac{\lambda^2+s^2}{2\eta}&\mapsto z_0\,,\\
\frac{\lambda^2-s^2}{2\eta}&\mapsto z_1\,,\\
\frac{\lambda}{\eta}r\cos\theta&\mapsto z_2\,,\\
\frac{\lambda}{\eta}r\sin\theta\cos\phi&\mapsto z_3\,, \\
\frac{\lambda}{\eta}r\sin\theta\sin\phi&\mapsto z_4\,,
\end{aligned}
\end{align}
\begin{figure*}[!htb]
\vspace*{-8mm}
\includegraphics[width=\textwidth]{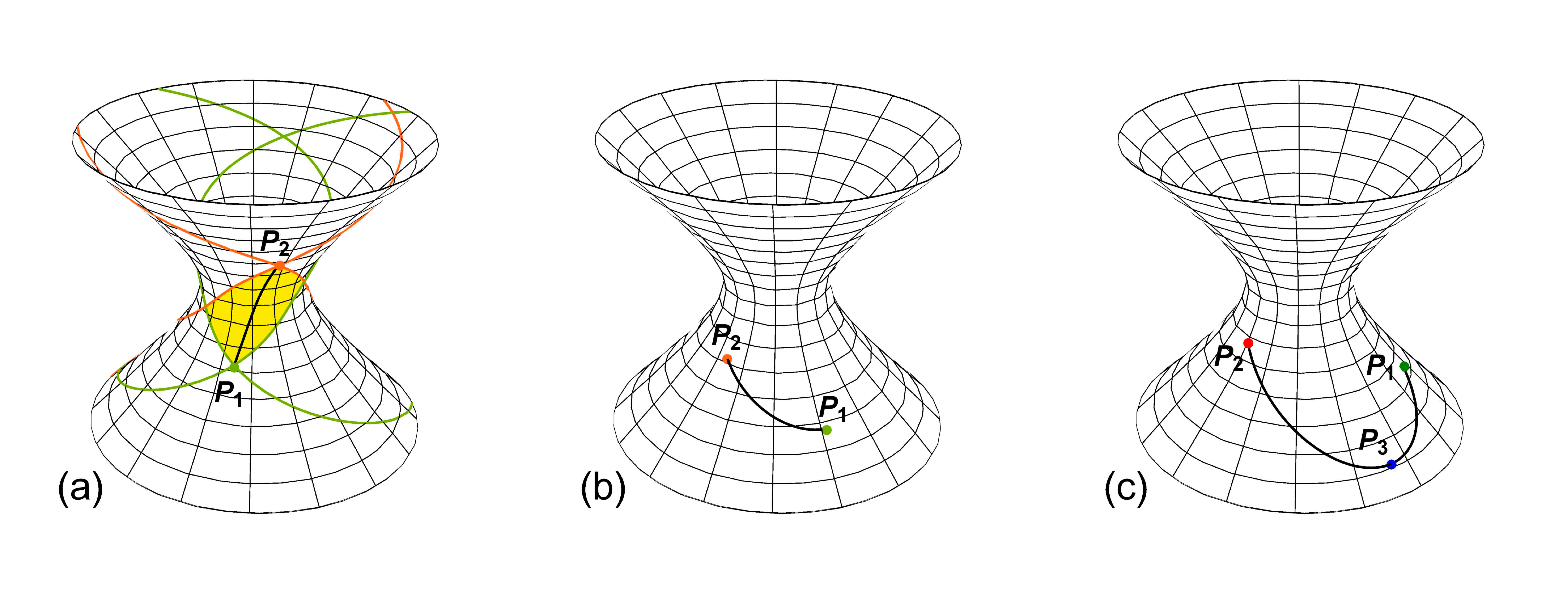}
\centering
\vspace*{-10mm}
\caption{\textbf{Geodesics on the 1+1 de Sitter manifold.} There are three classes of non-null geodesics on the de Sitter manifold. In (a), we see a future-directed timelike geodesic emanating from $P_1$ and terminating at $P_2$. These geodesics map out physical trajectories of subluminal objects within spacetime because the two points lie within each other's light cones, shown by the green and red lines. The Alexandroff set of points causally following $P_1$ and preceding $P_2$ is shown in yellow.
A spacelike geodesic joining two points with no causal overlap, shown in (b), ``bends away from the origin,'' meaning that in the plane defined by the origin of $\mathbb{M}^3$ and points $P_1,P_2$, this geodesic is farther from the origin than the Euclidean geodesic between the same points.
}
\label{fig1}
\end{figure*}
where $s^2\equiv r^2-\eta^2$, and the conformal time $\eta$ is defined as
\begin{equation}
\label{eq:conformal_time}
\eta(t)=\int^t\!\frac{dt'}{a(t')}\,.
\end{equation}
This embedding is a particular instance of the fact that any analytic $n$-dimensional pseudo-Riemannian manifold may be isometrically embedded into (at most) a $(n(n+1)/2)$-dimensional pseudo-Euclidean manifold (i.e., a flat metric with arbitrary non-Riemannian signature)~\cite{ref:friedman1961,ref:friedman1965}. The minimal $(n+1)$-dimensional embedding is most easily obtained using group theory by recognizing that the Lorentz group SO(1,3) is a stable subgroup of the de Sitter group dS(1,4) while the pseudo-orthogonal group SO(1,4) acts as its group of motions, i.e., dS(1,4)~=~SO(1,4)/SO(1,3), thereby indicating the minimal embedding is into the $\mathbb{M}^5$ space~\cite{ref:aldrovandi1995}. \par

If a spacelike geodesic extends far enough, there will exist an extremum, identified as $P_3$ in (c). As a result, it is simplest to use the spatial distance $\omega$ to parametrize these geodesics, though time can be used as well so long as those geodesics with turning points are broken into two parts at the point $P_3$.

Furthermore, it can be shown that geodesics on a de Sitter manifold follow the lines defined by the intersection of the hyperboloid with a hyperplane in $\mathbb{M}^5$ containing the origin and both endpoints of the geodesic~\cite{ref:asmus2009}. An illustration of both timelike and spacelike geodesics constructed this way in d$\mathbb{S}^2$ embedded in $\mathbb{M}^3$ can be found in Fig.~\ref{fig1}. This construction implies that the geodesic distance $d(x,y)$ in d$\mathbb{S}^4$ between two points $x$ and $y$ can be found using their inner product $\langle x,y\rangle=-x_0y_0+x_1y_1+x_2y_2+x_3y_3+x_4y_4$ in $\mathbb{M}^5$ via the following expression:
\begin{equation}
\label{eq:ds_5d}
d(x,y) = \begin{cases}\lambda\arccosh\frac{\langle x,y\rangle}{\lambda^2}&\text{if } x - y \text{ is timelike,} \\
0&\text{if } x - y \text{ is lightlike,} \\
\infty&\text{if } \langle x,y\rangle\leq -\lambda^2\text{ and } x\neq -y\,, \\
\lambda\arccos\frac{\langle x,y\rangle}{\lambda^2}&\text{otherwise.} \end{cases}
\end{equation}
\\
While there are many ways to find geodesic distances on a de Sitter manifold, this is perhaps the simplest one.

\section{The Geodesic Equations in Four Dimensions}
\label{sec:exact_solutions}
While de Sitter symmetries cannot be exploited in a general FLRW spacetime, it is still possible to solve the geodesic equations. A geodesic is defined in general by the variational equation
\begin{equation}
\delta\int ds = 0\,,
\end{equation}
which, if parametrized by parameter $\sigma$ ranging between two points $\sigma_1$ and $\sigma_2\,,$ becomes
\begin{equation}
\label{eq:geodesic_variation}
\delta\int_{\sigma_1}^{\sigma_2}\!\sqrt{g_{\mu\nu}\frac{\partial x^\mu}{\partial\sigma}\frac{\partial x^\nu}{\partial\sigma}}\,d\sigma = 0\,.
\end{equation}
The corresponding Euler-Lagrange equations obtained via the variational principle yield the well-known geodesic differential equations:
\begin{equation}
\label{eq:geodesic_diff_eq}
\nabla_X\frac{\partial x^\mu}{\partial\sigma} = \frac{\partial^2x^\mu}{\partial\sigma^2} + \Gamma_{\rho\tau}^\mu\frac{\partial x^\rho}{\partial\sigma}\frac{\partial x^\tau}{\partial\sigma} = \gamma\left(\sigma\right)\frac{\partial x^\mu}{\partial\sigma}\,,
\end{equation}
for the geodesic path $x^\mu(\sigma)$ with some as yet unknown function $\gamma(\sigma)$, where $\Gamma_{\rho\tau}^\mu$ are the Christoffel symbols defined by
\begin{equation}
\Gamma_{\rho\tau}^\mu = \frac{1}{2}g^{\mu\nu}\left(\frac{\partial g_{\nu\rho}}{\partial x^\tau} + \frac{\partial g_{\nu\tau}}{\partial x^\rho} - \frac{\partial g_{\rho\tau}}{\partial x^\nu}\right)\,,
\end{equation}
and $\nabla_X$ indicates the covariant derivative with respect to the tangent vector field $X$~\cite{ref:oneill1983}. If the parameter $\sigma$ is affine, then $\gamma(\sigma)=0$. To solve a particular problem with constraints, we must use both~\eqref{eq:geodesic_variation} and~\eqref{eq:geodesic_diff_eq}.

\subsection{The Differential Form of the Geodesic Equations}
If only the non-zero Christoffel symbols are kept, then \eqref{eq:geodesic_diff_eq} can be broken into two differential equations written in terms of the scale factor:
\begin{subequations}
\label{eq:geodesic2}
\begin{align}
\frac{\partial^2t}{\partial\sigma^2} + a\frac{da}{dt}h_{ij}\frac{\partial x^i}{\partial\sigma}\frac{\partial x^j}{\partial\sigma} &= \gamma\frac{\partial t}{\partial\sigma}\label{eq:geodesic2a}\,, \\
\frac{\partial^2x^i}{\partial\sigma^2}+\frac{2}{a}\frac{da}{dt}\frac{\partial t}{\partial\sigma}\frac{\partial x^i}{\partial\sigma} + \Gamma_{jk}^i\frac{\partial x^j}{\partial\sigma}\frac{\partial x^k}{\partial\sigma} &= \gamma\frac{\partial x^i}{\partial\sigma}\,, \label{eq:geodesic2b}
\end{align}
\end{subequations}
where $h_{ij}$ is the first fundamental form, i.e., the induced metric on a constant-time hypersurface, and the Latin indices are restricted to $\{1,2,3\}$. \par

To solve these, consider the spatial (Euclidean) distance $\omega$ between two points $\sigma_1$ and $\sigma_2$:
\begin{align}
\omega &= \int_{\sigma_1}^{\sigma_2}\!\sqrt{h_{ij}\frac{\partial x^i}{\partial\sigma}\frac{\partial x^j}{\partial\sigma}}\,d\sigma\,, \nonumber \\
\left(\frac{\partial\omega}{\partial\sigma}\right)^2 &= h_{ij}\frac{\partial x^i}{\partial\sigma}\frac{\partial x^j}{\partial\sigma}\,. \label{eq:omega_def}
\end{align}
This relation implies the spatial coordinates obey a geodesic equation with respect to the induced metric $h_{ij}$. Now,~\eqref{eq:geodesic2a} may be written in terms of $\omega$ using~\eqref{eq:omega_def}. The transformation needed for~\eqref{eq:geodesic2b} is found by multiplying by $h_{ij}(\partial x^j/\partial\sigma)$ and substituting the derivative of~\eqref{eq:omega_def} with respect to $\omega$:
\begin{align}
\begin{aligned}
\frac{\partial\omega}{\partial\sigma}\frac{\partial^2\omega}{\partial\sigma^2} &- \frac{1}{2}\frac{\partial h_{ij}}{\partial x^k}\frac{\partial x^i}{\partial\sigma}\frac{\partial x^j}{\partial\sigma}\frac{\partial x^k}{\partial\sigma} + \frac{2}{a}\frac{da}{dt}\frac{\partial t}{\partial\sigma}\left(\frac{\partial\omega}{\partial\sigma}\right)^2 \\
&+ h_{ij}\Gamma_{kl}^i\frac{\partial x^j}{\partial\sigma}\frac{\partial x^k}{\partial\sigma}\frac{\partial x^l}{\partial\sigma} = \gamma\left(\frac{\partial\omega}{\partial\sigma}\right)^2\,.
\end{aligned}
\end{align}
The second and fourth terms cancel by symmetry and, supposing $(\partial\omega/\partial\sigma)\neq 0$, the pair of equations~\eqref{eq:geodesic2} may be written as
\begin{subequations}
\label{eq:geodesic3}
\begin{align}
\frac{\partial^2 t}{\partial\sigma^2}+a\frac{da}{dt}\left(\frac{\partial\omega}{\partial\sigma}\right)^2 &= \gamma\frac{\partial t}{\partial\sigma}\,, \label{eq:geodesic3a} \\
\frac{\partial^2 \omega}{\partial\sigma^2} + \frac{2}{a}\frac{da}{dt}\frac{\partial t}{\partial\sigma}\frac{\partial\omega}{\partial\sigma} &= \gamma\frac{\partial\omega}{\partial\sigma}\,. \label{eq:geodesic3b}
\end{align}
\end{subequations}
We now proceed by parametrizing the geodesic by the Euclidean spatial distance, i.e., $\sigma\equiv\omega$. This yields $\partial\omega/\partial\sigma=1$, $\partial^2\omega/\partial\sigma^2=0$, and then~\eqref{eq:geodesic3b} gives $\gamma=(2/a)(da/dt)(\partial t/\partial\omega)$. Using these new relations,~\eqref{eq:geodesic3a} can be written as
\begin{equation}
\label{eq:geodesic3.5}
\frac{\partial^2 t}{\partial\omega^2} + \frac{da}{dt}\left(a - \frac{2}{a}\left(\frac{\partial t}{\partial\omega}\right)^2\right) = 0\,.
\end{equation}
While neither the spatial distance $\omega$ nor time $t$ are affine parameters along all Lorentzian geodesics, the results will not be affected, since the differential equations no longer refer to $\gamma$. We can see that if $\partial t/\partial\omega=0$ then the second derivative of $t$ is always negative for $t>0$ and positive for $t<0$, since $da/dt>0$ for expanding spacetimes:
\begin{equation}
\frac{\partial^2 t}{\partial\omega^2}=-a\frac{da}{dt}\,.
\end{equation}
If there exists a critical point exactly at $t=0$, it is a saddle point. From these facts, we conclude that any extremum found along a geodesic on a Friedmann-Lema\^itre-Robertson-Walker manifold is a local maximum in $t>0$ and a local minimum in $t<0$ with respect to $\omega$.\footnote{This statement is true under the assumption that the scale factor is a well-behaved monotonic function. If this condition does not hold, the following analysis must be reinspected.} An example of such a curve with an extremum is shown in Fig.~\ref{fig1}(c). \par

The second-order equation~\eqref{eq:geodesic3.5} may be simplified by multiplying by $2a^{-4}(\partial t/\partial\omega)$ and integrating by parts to get a non-linear first-order differential equation and a constant of integration $\mu$:
\begin{align}
0 &= \frac{\partial}{\partial\omega}\left[a^{-4}\left(\left(\frac{\partial t}{\partial\omega}\right)^2-a^2\right)\right]\,, \\
\frac{\partial\omega}{\partial t} &=\pm\left(a^2\left(t\right) + \mu a^4\left(t\right)\right)^{-1/2} \equiv G(t;\mu)
\,, \label{eq:geodesic4}
\end{align}
the right hand side of which is hereafter referred to as the geodesic kernel $G(t;\mu)$. We may neglect the sign by noting that the spatial distance $\omega$ should always be an increasing function of $t$, so that any integration of the geodesic kernel should be always be performed from past to future times. It will prove necessary to know the value of $\mu$ to find the final value of the geodesic length between two events.
\par

\subsection{The Integral Form of the Geodesic Equations}
To find the geodesic distance between a given pair of points/events, we need to use~\eqref{eq:geodesic4} in conjunction with the integral form of the geodesic equation, given in~\eqref{eq:geodesic_variation}. We begin by defining the integrand in~\eqref{eq:geodesic_variation} as the distance kernel $D(\sigma)$:
\begin{align}
D\left(\sigma\right) &\equiv \frac{ds}{d\sigma} = \sqrt{g_\munu\frac{\partial x^\mu}{\partial\sigma}\frac{\partial x^\nu}{\partial\sigma}}\,, \\
\intertext{so that the geodesic distance is}
d&=\int_{\sigma_1}^{\sigma_2}\!D\left(\sigma\right)\,d\sigma\,. \label{eq:distance_general}
\end{align}
The invariant interval~\eqref{eq:invariant_interval} tells us that $D^2$ is negative for timelike-separated pairs and positive for spacelike-separated ones, assuming $\sigma$ is monotonically increasing along the geodesic. Therefore, we always take the absolute value of $D^2$ so that the distance kernel is real-valued, while keeping in mind which type of geodesic we are discussing. \par

Depending on the particular scale factor and boundary values, we might sometimes parametrize the system using the spatial distance and other times using time. If we parametrize the geodesic with the spatial distance we find
\begin{align}
D\left(\omega\right) &= \sqrt{-\left(\frac{\partial t}{\partial\omega}\right)^2+a^2\left(t\left(\omega\right)\right)}\,, \label{eq:distances_general_spatial} \\
\intertext{and if we instead use time we get}
D\left(t\right) &= \sqrt{-1+a^2\left(t\right)\left(\frac{\partial\omega}{\partial t}\right)^2}\,, \label{eq:distances_general}
\end{align}
where the function $t(\omega)$ in the former equation is the inverted solution $\omega(t)$ to the differential equation~\eqref{eq:geodesic4}. Since the distance kernel is a function of the geodesic kernel, we will need to know the value $\mu$ associated with a particular set of constraints. \par

If we insert~\eqref{eq:geodesic4} into~\eqref{eq:distances_general}, we can can see what values the constant $\mu$ can take:
\begin{equation}
D\left(t\right) = \sqrt{\frac{-\mu a^2\left(t\right)}{1+\mu a^2\left(t\right)}}\,.
\end{equation}
If $D^2<0$ for timelike intervals, then $\mu>0$. If $\mu=0$, we obtain a lightlike geodesic, since the distance kernel becomes zero. Hence, spacelike intervals correspond to $-a^{-2}(t)<\mu<0$. We do not consider $\mu<-a^{-2}(t)$ because this corresponds to an imaginary $\partial\omega/\partial t$, which we consider non-physical.

\par

\subsection{Geodesic Constraints and Critical Points}
We would like to find geodesics for both initial-value and boundary-value problems. If we have Cauchy boundary conditions, i.e., the initial position and velocity vector are known, then finding $\mu$ is simple: since the left hand side of~\eqref{eq:geodesic4} is just the speed $v_0\equiv|v^i(t_0)|$, where $v^i$ is the velocity vector defined by our initial conditions, we have
\begin{equation}
\label{eq:cauchy}
\mu=a_0^{-2}\left(v_0^{-2}a_0^{-2}-1\right)\,,
\end{equation}
where $a_0\equiv a(t_0)$. This allows for simple solutions to cases with Cauchy
boundary conditions.

However, if we have Dirichlet boundary conditions, i.e., the initial and final positions are known, then we must integrate~\eqref{eq:geodesic4} instead. The bounds of such an integral need to be carefully considered: if we have a spacelike geodesic which starts and ends at the same time, for instance, then it is not obvious how to integrate the geodesic kernel. In fact, we face an issue with the boundaries whenever we have geodesics with turning points. This feature occurs whenever $\partial t/\partial\omega = 0$, i.e.,
\begin{equation}
\label{eq:max_time_constraint}
a(t_c) = \pm\sqrt{-\mu^{-1}}\,.
\end{equation}
Since in this case $\mu < 0$, we see that turning points only occur for spacelike
geodesics. Furthermore, since all of the scale factors given by~\eqref{eq:scale_factors} are monotonic, this situation occurs in such spacetime only at a single point along a geodesic, if at all, identified as $P_3$ in Fig.~\ref{fig1}(c). Specifically, if $t_1,t_2,t_3$ respectively correspond to the times at points $P_1,P_2,P_3$, then the integral of the geodesic kernel is found by integrating from $t_3$ to $t_1$ as well as from $t_3$ to $t_2$, since time is not monotonic along the geodesic. If no such turning point $P_3$ exists along the geodesic, a single integral from $t_1$ to $t_2$ may be performed. The integral of the distance kernel should be performed in the same way for the same reasons. \par

To determine if a turning point exists along a spacelike geodesic, we begin by noting that there is a corresponding critical spatial distance $\omega_c$ which corresponds to the critical time defined in~\eqref{eq:max_time_constraint}. If we suppose $t_2>t_1>0$, then the geodesic kernel is maximized when $\mu=\mu_c\equiv-a^{-2}(t_2)$, i.e., when $\mu$ attains its minimum value. This is the minimum value of $\mu$ along the geodesic, since
$a(t)$ is monotonically increasing. The critical spatial distance is defined by this
$\mu_c$ and is given by
\begin{equation}
\label{eq:critical_separation}
\omega_c = \int_{t_1}^{t_2}\!\frac{1}{a(t)}\left(1-\left(\frac{a(t)}{a(t_2)}\right)^2\right)^{-1/2}\,dt\,.
\end{equation}
Since $\mu_c$ maximizes the geodesic kernel, it is impossible for a spacelike-separated pair to be spatially farther apart without their geodesic having a turning point. We then conclude that if $\omega<\omega_c$ for a particular pair of spacelike-separated points, then the geodesic is of the form shown in Fig.~\ref{fig1}(b), and if $\omega>\omega_c$ it is of the form shown in Fig.~\ref{fig1}(c). In other words, if the geodesic is of the latter type, then the solution to~\eqref{eq:geodesic4} is
\begin{align}
\omega&=\int_{t_1}^{t_c}\!G\left(t;\mu\right)\,dt + \int_{t_2}^{t_c}\!G\left(t;\mu\right)\,dt\,,\label{eq:omega_coupled} \\
\intertext{while the solution to~\eqref{eq:distance_general} using~\eqref{eq:distances_general} is}
d&=\int_{t_1}^{t_c}\!D\left(t\right)\,dt+\int_{t_2}^{t_c}\!D\left(t\right)\,dt\,,
\end{align}
again supposing $t_2>t_1>0$. The bounds on the integral are chosen this way due to
the change of sign in the geodesic kernel on opposite sides of the critical point.
If $0>t_2>t_1$ then the bounds on the integrals are reversed so that $\omega,d>0$. \par

\begin{figure*}[!htb]
\includegraphics[width=\textwidth]{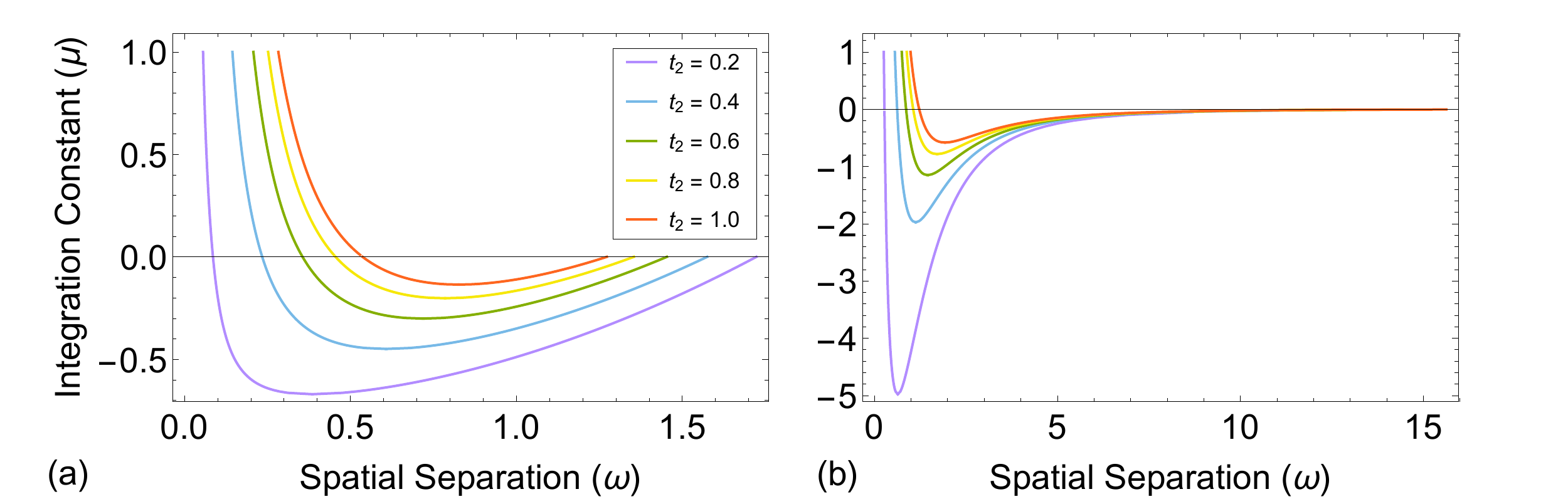}
\centering
\caption{\textbf{Evidence of geodesic horizons in FLRW manifolds.} Certain FLRW manifolds are not spacelike-geodesically-connected, such as the de Sitter manifold. In (a) we see the relation between the integration constant $\mu$, first defined in~\eqref{eq:geodesic4}, and the spatial separation between two points on the de Sitter manifold. The initial point is located at $t_1=0.1$ and the curves show the behavior for several choices of the final time $t_2$. For small $\omega$, the pair of points is timelike-separated and $\mu$ is positive. As $\omega$ tends to zero, $\mu$ tends to infinity, indicating the manifold is timelike-geodesically-complete. As $\omega$ increases and the geodesic becomes spacelike, it will ultimately have a turning point at $\omega_c$, located at the minimum of each curve and defined by~\eqref{eq:critical_separation}. Ultimately, for manifolds which are spacelike-geodesically-incomplete the curve terminates at some maximum spatial separation $\omega_m$ defined by~\eqref{eq:maximum_separation}. In (b), showing the Einstein-de Sitter manifold case, the curves extend to infinity on the right because the manifold is geodesically complete.}
\label{fig2}
\end{figure*}

\subsection{Geodesic Connectedness}
\label{sec:geo_conectedness}
Certain FLRW manifolds are not spacelike-geodesically-connected, meaning not all pairs of spacelike-separated points are connected by a geodesic. For a given pair of times $t_1,t_2$ there exists a maximum spatial separation $\omega_m$ past which the two points cannot be connected by a geodesic. To determine this maximum spatial distance $\omega_m$ for a particular pair of points, we use~\eqref{eq:omega_coupled}, this time taking the limit $\mu\to 0^-$. This limit describes a spacelike geodesic which is asymptotically becoming lightlike. If the critical time $t_c$ remains finite in this limit, the manifold is geodesically connected and $\omega_m=\infty$, whereas if it becomes infinite then $\omega_m$ remains finite, shown in detail in Fig.~\ref{fig2}. The equation~\eqref{eq:omega_coupled} in the limit $\mu\to 0^-$ is
\begin{equation}
\label{eq:maximum_separation}
\omega_m = \int_{t_1}^{t_c}\!\frac{dt}{a(t)} + \int_{t_2}^{t_c}\!\frac{dt}{a(t)}\,.
\end{equation}
Comparing~\eqref{eq:maximum_separation} to~\eqref{eq:conformal_time} we notice that $\omega_m$ is simply a combination of conformal times using the boundary points $t_1$ and $t_2$: $\omega_m\propto\eta_c\equiv\eta(t_c)$, and so if $\eta_c$ is finite, then $\omega_m$ will be finite as well. Therefore, we conclude that a Friedmann-Lema\^itre-Robertson-Walker manifold is geodesically complete if
\begin{align}
\lim_{\mu\to 0^-}\left|\eta_c\right|&=\infty\,, \label{eq:geo_compl}\\
\intertext{where $\eta_c$ is obtained by inverting}
a(t(\eta_c))&=\pm\sqrt{-\mu^{-1}}\,,
\end{align}
using the appropriate $a(t)$ and $t(\eta)$ for the given manifold. \par

As an example, consider the de Sitter manifold:
\begin{equation}
\frac{\lambda}{\eta_c} = \pm\sqrt{-\mu^{-1}}\,,
\end{equation}
so that the limit maximum conformal time in terms of $\mu$ is
\begin{equation}
\lim_{\mu\to 0^-}\left|\eta_c\right| = \lim_{\mu\to 0^-}\lambda\sqrt{-\mu} = 0\,.
\end{equation}
Therefore, in the flat foliation, there exist pairs of points on the de Sitter manifold which cannot be connected by a geodesic. On the other hand, if we consider the Einstein-de Sitter manifold, which represents a spacetime with dust matter, the scale factor is proportional to $\eta^2$:
\begin{equation}
\lim_{\mu\to 0^-}\left|\eta_c\right| \propto \lim_{\mu\to 0^-}\left(-\mu^{-1}\right)^{1/4} = \infty\,,
\end{equation}
so that every pair of points may be connected by a geodesic.

\section{Examples}
\label{sec:examples}
Here we apply the results above to calculate geodesics in the FLRW manifolds defined by each of the scale factors in~\eqref{eq:scale_factors}, using two type of constraints: the Dirichlet and Cauchy boundary conditions. The former conditions specify two events or points in a given spacetime that can be either timelike or spacelike separated, as in Fig.~\ref{fig1}. The latter conditions specify just one point and a vector of initial velocity. If the initial speed is below the speed of light, then the resulting geodesic is timelike, and corresponds to a possible world line of a massive particle. If the initial speed is above the speed of light, i.e., the initial tangent vector is spacelike, then the resulting geodesic is spacelike, and corresponds to a geodesic of a hypothetical superluminal particle. Even though tachyons may not exist, spacelike geodesics are well defined mathematically.
The last example that we consider illustrates how to apply these techniques to find numerical values for geodesic distances in our physical universe.
\subsection{Dark Energy}
Suppose we wish to find the geodesic distance using the Dirichlet boundary conditions $\{t_1,\ab t_2,\ab\omega\}$. The geodesic kernel in a flat de Sitter spacetime is
\begin{equation}
G_\Lambda\left(t;\mu\right) = \lambda^{-1}\left(e^{2t/\lambda}+\mu e^{4t/\lambda}\right)^{-1/2}\,,
\end{equation}
where $\mu$ has absorbed a factor of $\lambda^2$ and we use $\eta\in[-1,0)$ so that $t\geq0$. We can easily transform the kernel into a polynomial equation by using the conformal time:
\begin{equation}
G_\Lambda\left(\eta;\mu\right) = \left(1+\frac{\mu}{\eta^2}\right)^{-1/2}\,.
\end{equation}
If the minimal value of $\mu$ is inserted into this kernel, the turning point $\omega_c$ can be found exactly:
\begin{align}
\mu_c &= -\eta_2^2\,, \\
G_\Lambda\left(\eta;\mu_c\right) &= \left(1-\left(\frac{\eta_2}{\eta}\right)^2\right)^{-1/2}\,, \\
\omega_c\left(\eta_1,\eta_2;\mu_c\right) &= \int_{\eta_1}^{\eta_2}\!G_\Lambda\left(\eta;\mu_c\right)\,d\eta\,, \nonumber \\
  &= \sqrt{\eta_1^2-\eta_2^2}\,.
\end{align}
The geodesic kernel may now be integrated both above and below the turning point:
\begin{equation}
\label{eq:ds_mu}
\omega = \begin{cases}
\sqrt{\eta_1^2+\mu}-\sqrt{\eta_2^2+\mu}&\text{if } \omega<\omega_c\,, \\
\sqrt{\eta_1^2+\mu}+\sqrt{\eta_2^2+\mu}&\text{if } \omega>\omega_c\,.
\end{cases}
\end{equation}
The variable $\mu$ is then found by inverting one of these equations. Finally, substitution of the scale factor and numerical value $\mu$ into~\eqref{eq:distances_general} gives the geodesic distance for a pair of coordinates defined by $\{\eta_1,\ab \eta_2,\ab\omega\}$:
\begin{subequations}
\label{eq:ds_4d}
\begin{align}
&d_\Lambda\left(t_1,t_2;\mu\right) = \sinh^{-1}\left(\frac{\sqrt\mu}{\eta_1}\right)-\sinh^{-1}\left(\frac{\sqrt\mu}{\eta_2}\right)\,,\\
\intertext{for timelike-separated pairs, and}
\begin{split}
&d_\Lambda\left(t_1,t_2;\mu\right) =\\
&\begin{cases}
\sinh^{-1}\left(\frac{\sqrt{-\mu}}{\eta_1}\right)-\sinh^{-1}\left(\frac{\sqrt{-\mu}}{\eta_2}\right)&\text{if } \omega<\omega_c\,, \\
\sinh^{-1}\left(\frac{\sqrt{-\mu}}{\eta_1}\right)+\sinh^{-1}\left(\frac{\sqrt{-\mu}}{\eta_2}\right)+\pi&\text{if } \omega>\omega_c\,,
\end{cases}
\end{split}
\end{align}
\end{subequations}
for spacelike-separated pairs. In Appendix~\ref{sec:equivalence} we show that this solution is equivalent to the solution found using the embedding in Sec.~\ref{sec:desitter}.

\subsection{Dust}
\label{sec:dust}
In this example, let us suppose we have Cauchy boundary conditions and we want an expression for the geodesic distance in terms of spatial distance traveled $\omega$. First, knowing the values $(t_0,r_0,\theta_0,\phi_0)$ and $|v_0|$, we can find the parameter $\mu$ via~\eqref{eq:cauchy}. Because the manifold has a
singularity at $t=0$, we assert $t_0\neq0$ to avoid a nonsensical value for $\mu$. We proceed by parametrizing the geodesic equation by the spatial distance, following~\eqref{eq:distances_general_spatial}, so that the distance kernel for this spacetime is
\begin{equation}
D_D\left(\omega\right) = \alpha^2\left|\mu\right|^{1/2}\left(\frac{3t\left(\omega\right)}{2\lambda}\right)^{4/3}\,.
\end{equation}
We use the geodesic kernel to find $t(\omega)$ directly, by solving~\eqref{eq:geodesic4} for $\omega(t)$ and inverting the solution. In the spacetime with dust matter and no cosmological constant the geodesic kernel is
\begin{equation}
G_D\left(t;\mu\right) = \left(\alpha^2\left(\frac{3t}{2\lambda}\right)^{4/3}+\mu\alpha^4\left(\frac{3t}{2\lambda}\right)^{8/3}\right)^{-1/2}\,,
\end{equation}
\\
\noindent which, using the transformations $x\equiv(3t/2\lambda)^{1/3}$ and $\mu\to\alpha^2\mu$, becomes
\begin{equation}
G_D\left(x;\mu\right) = \frac{2\lambda}{\alpha}\left(1+\mu x^4\right)^{-1/2}\,.
\end{equation}
The value of $\omega$ where the turning point occurs is then
\begin{equation}
\begin{split}
\omega_c(x_0;\mu) = &\frac{2\lambda}{\alpha}\left(\frac{\sqrt{\pi}\,\Gamma(5/4)}{\Gamma(3/4)}\left(-\mu\right)^{-1/4}\right. \\
&\left.- x_0\,{}_2F_1\left(\frac{1}{4},\frac{1}{2};\frac{5}{4};-\mu x_0^4\right)\right)\,,
\end{split}
\end{equation}
where $x_0\equiv x(t_0)$ and ${}_2F_1(a,b;c;z)$ is the Gauss hypergeometric function. \par

The final expression $\omega(t)$ still depends on the existence of a critical point along the geodesic. To demonstrate how piecewise solutions are found, hereafter we suppose we are studying a superluminal inertial object moving fast and long enough to take a geodesic with a turning point. The spatial distance $\omega(x;x_0)$, with $x>x_0$ and $\mu<0$, which we know because the geodesic is spacelike, is
\begin{subequations}
\begin{align}
\begin{split}
&\omega^{(1)}\left(x;x_0,x_c\right) = \frac{2\lambda}{\alpha}x_c\left(F\left(\arcsin\left(\frac{x}{x_c}\right)\bigg|-1\right)\right. \\
&\left.\qquad\qquad\qquad\quad-F\left(\arcsin\left(\frac{x_0}{x_c}\right)\bigg|-1\right)\right)\,,
\end{split}\\
\intertext{before the critical point, and}
\begin{split}
&\omega^{(2)}\left(x;x_0,x_c\right) = \frac{2\lambda}{\alpha}x_c\bigg(2K\left(-1\right) \\
&- F\left(\arcsin\left(\frac{x_0}{x_c}\right)\bigg|-1\right) - F\left(\arcsin\left(\frac{x}{x_c}\right)\bigg|-1\right)\bigg)\,,
\end{split}
\end{align}
\end{subequations}
afterward, where $x_c=(-\mu)^{-1/4}$, and $K(m)$ and $F(\phi|m)$ respectively are the complete and incomplete elliptic integrals of the first kind with parameter $m$. These expressions $\omega(x;x_0,x_c)$ are slightly different for $\mu>0$. Despite the apparent complexity of the above expressions, they are in fact easy to invert via the Jacobi elliptic functions. The distance for a geodesic with a turning point is
\begin{equation}
d\left(\omega;\mu\right) = \int_0^{\omega_c}\!D\left(\omega^{(1)}\right)\,d\omega + \int_{\omega_c}^{\omega}\!D\left(\omega^{(2)}\right)\,d\omega\,,
\end{equation}
giving the final result
\begin{equation}
\begin{split}
&d_D\left(\omega;\mu\right) = \frac{\alpha^2\left|\mu\right|^{1/2}x_c^4}{3\beta_1}\Bigg(\beta_1\left(2\omega_c-\omega\right) \\
&+ \frac{x_0}{x_c}\sqrt{1-\left(\frac{x_0}{x_c}\right)^4} + \fn\left(\beta_3-\beta_1\omega_c|-1\right) \\
&- \fn\left(\beta_1\omega_c+\beta_2|-1\right) - \fn\left(\beta_1\omega-\beta_3|-1\right)\Bigg)\,,
\end{split}
\end{equation}
where we have used the auxiliary variables
\begin{align}
\beta_1 &\equiv \frac{\alpha}{2\lambda x_c}\,, \\
\beta_2 &\equiv F\left(\arcsin\left(\frac{x_0}{x_c}\right)\bigg|-1\right)\,, \\
\beta_3 &\equiv 2K\left(-1\right) - \beta_2\,, \\
\fn\left(\phi|m\right) &\equiv \sn\left(\phi|m\right)\cn\left(\phi|m\right)\dn\left(\phi|m\right)\,,
\end{align}
and the three functions in the last definition are the Jacobi elliptic functions with parameter $m$.

\subsection{Radiation}
Here we suppose we have Cauchy boundary conditions, but the particle will take a timelike geodesic, i.e., $\mu>0$. Using the transformations $x\equiv\sqrt{2t/\lambda}$ and $\mu\to\alpha^{3/2}\mu$, we can write the geodesic kernel as
\begin{equation}
G_R\left(x;\mu\right) = \frac{\lambda}{\alpha^{3/4}}\left(1+\mu x^2\right)^{-1/2}\,.
\end{equation}
If this kernel is integrated over $x$ to find the spatial distance $\omega(x)$, the result can be inverted to give
\begin{equation}
x\left(\omega;\mu,x_0\right) = \mu^{-1/2}\sinh\left(\beta_1\omega+\beta_2\right)\,,
\end{equation}
where $\beta_1\equiv\alpha^{3/4}\mu^{1/2}/\lambda$ and $\beta_2\equiv\arcsinh(\mu^{1/2}x_0)$. Since the geodesic distance is more easily found when we parametrize with the spatial distance $\omega$, we can write the distance kernel as
\begin{equation}
D_R\left(\omega\right) = \alpha^{3/2}\mu^{1/2}x^2\left(\omega;\mu,x_0\right)\,,
\end{equation}
and the geodesic distance as
\begin{align}
d_R\left(\omega;\mu,x_0\right) &=\int_0^\omega\!D_R\left(\omega^\prime\right)\,d\omega^\prime\,, \\
&\begin{aligned}
= &\frac{\alpha^{3/2}}{4\mu^{1/2}\beta_1}\left(\sinh\left(2\left(\beta_1\omega+\beta_2\right)\right)\right.\\
 &\left.- \sinh\left(2\beta_2\right) - 2\beta_1\omega\right)\,.
\end{aligned}
\end{align}
Typically, timelike geodesics are parametrized by time: since there exists a closed-form solution for $\omega(x(t))$ this expression can be substituted here, though it would needlessly add extra calculations. Therefore, in practice it is computationally simpler to use a spatial parametrization.

\subsection{Stiff Fluid}
Suppose we have a spacetime containing a homogeneous stiff fluid, and we wish to find a timelike geodesic using Dirichlet boundary conditions. Using the transformation $x\equiv(3t/\lambda)^{1/3}$, we can write the geodesic kernel as
\begin{equation}
G_S\left(x;\mu\right) = \frac{\lambda}{\alpha^{1/2}}\frac{x}{\left(1+\mu x^2\right)^{1/2}}\,,
\end{equation}
where $\mu$ has absorbed a factor of $\alpha$. This kernel can easily be integrated to find
\begin{equation}
\omega\left(x_0,x_1;\mu\right) = \frac{\lambda}{\alpha^{1/2}\mu}\left(\sqrt{1+\mu x_1^2} - \sqrt{1+\mu x_0^2}\right)\,.
\end{equation}
\\
\noindent The constant $\mu$ may be found provided the initial conditions $\{x_0,x_1,\omega\}$:
\begin{equation}
\mu = \frac{x_0^2+x_1^2}{\omega^2} - 2\sqrt{\frac{x_0^2x_1^2 + \omega^2}{\omega^4}}\,.
\end{equation}
Finally, if the geodesic is parametrized by $x(t)$ we arrive at
\begin{equation}
\begin{split}
d_S(x_0,&x_1;\mu) = \frac{\lambda}{3\mu}\bigg(2\sqrt{x_1^2 + \mu^{-1}} - 2\sqrt{x_0^2 + \mu^{-1}} \\
& - x_1^3\sqrt{\mu\left(x_1^{-2} + \mu\right)} + x_0^3\sqrt{\mu\left(x_0^{-2} + \mu\right)}\bigg)\,.
\end{split}
\end{equation}

\subsection{Dark Energy and Dust}
\label{sec:mixed}
None of the spacetimes with a mixture of dark energy and some form of matter have closed-form solutions for geodesics, because the scale factors are various powers of the hyperbolic sine function, so it becomes cumbersome to work with the geodesic and distance kernels. However, by using the right transformations, it is still possible to make the problem well-suited for fast numerical integration. In this example, we use the mixed dust and dark energy spacetime, following the same procedure as before; for other spacetimes with mixed contents the same method applies. This time, the geodesic kernel is
\begin{equation}
G_{\Lambda D}\left(t;\mu\right) = \left(\sinh^{4/3}\left(\frac{3t}{2\lambda}\right)+\mu\sinh^{8/3}\left(\frac{3t}{2\lambda}\right)\right)^{-1/2}\,.
\end{equation}
Once again, the kernel can be written as a polynomial expression, this time using the square root of the scale factor as the transformation:
\begin{align}
x\left(t\right) &\equiv \sinh^{1/3}\left(\frac{3t}{2\lambda}\right)\,, \nonumber \\
G_{\Lambda D}\left(x;\mu\right) &= 2\left(\left(1+x^6\right)\left(1+\mu x^4\right)\right)^{-1/2}\,.
\end{align}
There is no known closed-form solution to the integral of $G_{\Lambda D}$. The distance kernel is best represented as a function of $t$ to simplify numerical evaluations:
\begin{equation}
D_{\Lambda D}\left(t\right) = \sqrt{\frac{-\mu\sinh^{2/3}\left(3t/\lambda\right)}{1+\mu\sinh^{2/3}\left(3t/\lambda\right)}}\,.
\end{equation}
There is no known closed-form solution to this kernel's integral either, but it can be quickly computed numerically, since the hyperbolic term needs to be evaluated only once for each value of $t$. In general, the numeric evaluations of such integrals can be quite fast if the kernels take a polynomial form, and a Gauss-Kronrod quadrature can be used for numeric evaluation of these integrals.

\subsection{Dark Energy, Dust, and Radiation}
Typically in cosmology one studies one particular era, whether the early
inflationary phase, the radiation-dominated phase, the matter-dominated phase
after recombination, or ultimately today's period of accelerated
expansion. Perhaps the most important spacetime which we have not looked at
yet is the FLRW spacetime which most closely models our own physical universe,
in its entirety. In this section we will show how to most efficiently find geodesics in our (FLRW $\Lambda$DR) universe. \par

Because the scale factor $a(t)$ is a smooth, monotonic, differentiable, and bijective function of time, it, instead of time $t$ or spatial distance $\omega$, can parametrize geodesics, so long as we remember to break up expressions when there exists a turning point in long spacelike geodesics. In what follows we will restrict the analysis to timelike geodesics for simplicity. To find spacelike geodesics, refer to the steps performed in Sec.~\ref{sec:dust}. Using the scale-factor parametrization, the geodesic and distance kernels are
\begin{align}
&G_{\Lambda DR}\left(a;\mu\right) = \lambda\left[\left(1+\mu a^2\right)\left(\frac{\Omega_R}{\Omega_\Lambda}+\frac{\Omega_D}{\Omega_\Lambda}a+a^4\right)\right]^{-1/2}\,, \label{eq:full_gkernel} \\
&D_{\Lambda DR}\left(a\right) = \lambda\left[\left(\frac{-\mu a^4}{1+\mu a^2}\right)\left(\frac{\Omega_R}{\Omega_\Lambda}+\frac{\Omega_D}{\Omega_\Lambda}a+a^4\right)^{-1}\right]^{-1/2}\,. \label{eq:full_dkernel}
\end{align}
As we saw in Sec.~\ref{sec:mixed}, integrands such as these produce no closed-form solutions, but they are easily evaluated numerically due to their polynomial form. \par

We now provide a simple example of computing an exact geodesic distance between a pair of events in our physical universe using these results. Suppose we are to measure the timelike geodesic distance between an event in the early universe, where $t_1=10^{11}$s, and another event near today, $t_2=4.3\times10^{17}$s. Let the spatial distance of this geodesic be $\omega=4.1\times10^{13}$km, roughly the distance to Alpha Centauri. Taking relevant experimental values from recent measurements~\cite{calabrese2017cosmological}, we find the Hubble constant is $H_0=100h$ km/s/Mpc, where $h=0.705$, and the density parameters are $\Omega_\Lambda=0.723$, $\Omega_D=0.277$, and $\Omega_R=9.29\times10^{-5}$. The leading constant $\lambda$ in the above equations can be expressed as $\lambda=H_0^{-1}\Omega_\Lambda^{-1/2}$, thereby completing the set of all the relevant physical parameters used in~\eqref{eq:full_gkernel} and~\eqref{eq:full_dkernel}.
We then integrate the geodesic kernel~\eqref{eq:full_gkernel}, inserting the speed of light $c$ where needed, to numerically solve for the integration constant $\mu$, which we find to be $\mu=2.53\times10^{23}$. Inserting this value into the distance kernel~\eqref{eq:full_dkernel} and evaluating numerically gives a final geodesic distance of $d=2.22\times10^{23}$km.

\section{Conclusion}
\label{sec:conclusion}
By integrating the geodesic differential equations~\eqref{eq:geodesic_diff_eq} we have shown for spacetimes with dark energy, dust, radiation, or a stiff fluid, that it is possible to find a closed-form solution for the geodesic distance provided either initial-value or boundary-value constraints. Furthermore, by studying the form of the first-order differential equation~\eqref{eq:geodesic4} we found that extrema along spacelike geodesic curves will always point away from the origin. This insight provides a better understanding of how to integrate the geodesic and distance kernels~(\ref{eq:geodesic4},~\ref{eq:distances_general_spatial},~\ref{eq:distances_general}) for different types of boundary conditions. Moreover, our other important result in Sec.~\ref{sec:geo_conectedness} demonstrates how, using~\eqref{eq:conformal_time},~\eqref{eq:max_time_constraint} and~\eqref{eq:geo_compl}, we are able to tell, using only the scale factor, whether or not all points on a flat FLRW manifold can be connected by a geodesic. This observation is particularly useful in numeric experiments and investigations that can study only a finite portion of a spatially flat manifold. Finally, in Sec.~\ref{sec:examples} we provided several examples of how these results might be applied to some of the most well-studied FLRW manifolds, including the manifold describing our universe. While not all spacetimes have closed-form solutions for geodesics, it is still possible to reframe the problem in a way which may be solved efficiently using numerical methods in existing software libraries.

\acknowledgments
We thank Cody Long, Aron Wall, and Michel Buck for useful discussions and suggestions. This work was supported by NSF grants No.\ CNS-1442999, CCF-1212778, and PHY-1620526, ARO grant No.\ W911NF-16-1-0391, and DARPA grant No.\ N66001-15-1-4064. Any opinions, findings, and conclusions or recommendations expressed in this publication are those of the authors and do not necessarily reflect the views of DARPA.

\appendix
\section*{Appendix: Equivalence\\ of de Sitter Solutions}
\label{sec:equivalence}
Here we show that the equations~\eqref{eq:ds_5d} and~\eqref{eq:ds_4d} are equal under certain assumptions. Let us refer to the former as $d_1$ and the latter as $d_2$. The conformal time in the de Sitter spacetime is $\eta(t) = -e^{-t/\lambda}$, with $\eta\in[-1,0)$ so that the cosmological time $t$ remains positive. Since the geodesic distance depends on the spatial distance, but not the individual spatial coordinates, we can assume without loss of generality that the initial point is located at the origin, $r=\theta=\phi=0$, and the second point is located at some distance $\omega$ from the origin, $r=\omega,\,\theta=\phi=0$. Further, to simplify the proof, suppose the initial point is at time $t=0$ ($\eta=-1$) and the second point at some $t=t_0>0$ ($\eta=\eta_0\in(-1,0)$). We are allowed to make these assumptions due to the spatial symmetries associated with the dS(1,3) group and the existence of a global timelike Killing vector in the flat foliation of the de Sitter manifold~\cite{ref:podolsky1993}. In addition, suppose the geodesic is timelike so that $\omega\in[0,\eta_0+1)\subseteq[0,1)$. This same method may be applied to spacelike geodesics. \par

Using these values, the embedding coordinates in $\mathbb{M}^5$ are $x=((1-\lambda^2)/2,\,-(1+\lambda^2)/2,\,0,\,0,\,0)$ and $y=((\lambda^2+\omega^2-\eta_0^2)/2\eta_0,\,(\lambda^2-\omega^2+\eta_0^2)/2\eta_0,\,\lambda\omega/\eta_0,\,0,\,0)$. This equation gives a geodesic distance
\begin{equation}
d_1 = \lambda\arccosh\left(\frac{\omega^2-\eta_0^2-1}{2\eta_0}\right)\,.
\end{equation}
On the other hand, we can use the solution provided by~\eqref{eq:ds_4d} using the value of $\mu$ in~\eqref{eq:ds_mu}:
\begin{equation}
\label{eq:app_mu}
\mu=\frac{\left(\omega+\eta_0+1\right)\left(\omega+\eta_0-1\right)\left(\omega-\eta_0+1\right)\left(\omega-\eta_0-1\right)}{4\lambda^2\omega^2}\,,
\end{equation}
\\
\noindent in the geodesic distance expression
\begin{equation}
d_2 = \lambda\left(\arcsinh\left(\frac{\lambda\sqrt{\mu}}{-\eta_0}\right) - \arcsinh\left(\lambda\sqrt{\mu}\right)\right)\,.
\end{equation}
If we apply $\cosh(d/\lambda)$ to each of these expressions, and use the identities $\cosh(x-y) = \cosh x\cosh y - \sinh x\sinh y$ and $\cosh\arcsinh x = \sqrt{x^2+1}$, we may equate them to get
\begin{equation}
\frac{\omega^2-\eta_0^2-1}{2\eta_0} = \sqrt{\left(\lambda^2\mu+1\right)\left(\frac{\lambda^2\mu}{\eta_0^2}+1\right)}+\frac{\lambda^2\mu}{\eta_0}\,.
\end{equation}
Using~\eqref{eq:app_mu} and some algebra, the right hand side may be simplified to give the result on the left hand side, thereby proving they are equal.

\bibliographystyle{apsrev4-1}
\bibliography{paper,geo}

\end{document}